\documentclass[a4paper]{article}

\usepackage{INTERSPEECH2019}
\usepackage{url}
\usepackage{algorithm}
\usepackage[noend]{algpseudocode}
\newcommand{\argmax}[1]{\underset{#1}{\operatorname{arg}\,\operatorname{max}}\;}
\usepackage{subfig}
\usepackage{cite}

\title{Learning Speaker Representations with Mutual Information}
\name{Mirco Ravanelli, Yoshua Bengio$^{*}$}
\address{
  Mila, Universit\'e de Montr\'eal , $^*$CIFAR Fellow}

\email{mirco.ravanelli@gmail.com}

\begin{document}
\ninept

\maketitle
\begin{abstract}
Learning good representations is of crucial importance in deep learning. 
Mutual Information (MI) or similar measures of statistical dependence are promising tools for learning these representations in an unsupervised way. Even though the mutual information between two random variables is hard to measure directly in high dimensional spaces, some recent studies have shown that an implicit optimization of MI can be achieved with an encoder-discriminator architecture similar to that of Generative Adversarial Networks (GANs).

In this work, we learn representations that capture speaker identities by maximizing the mutual information between the encoded representations of chunks of speech randomly sampled from the same sentence.
The proposed encoder relies on the SincNet architecture and transforms raw speech waveform into a compact feature vector. The discriminator is fed by either positive samples (of the joint distribution
of encoded chunks) or negative samples (from the product of the marginals) and is trained to separate them.

We report experiments showing that this approach effectively learns useful speaker representations, leading to promising results on speaker identification and verification tasks. Our experiments consider both unsupervised and semi-supervised settings and compare the performance achieved with different objective functions.
\end{abstract}

\section{Introduction}
\label{sec:intro}
Deep learning has shown remarkable success in numerous speech tasks, including speech recognition \cite{lideng,IEEEexample:intro1,ravanelli_thesis,ravanelli_icassp} and speaker recognition \cite{speaker_rec_dnn,dnn_speaker_rec2}.
The deep learning paradigm aims to describe data by means of a hierarchy of representations, that are progressively combined to model higher level abstractions \cite{Goodfellow-et-al-2016-Book}.
Most commonly, deep neural networks are trained in a supervised way, while learning meaningful representations in an unsupervised fashion is more challenging but could be useful especially in semi-supervised settings. 

Several approaches have been proposed for deep unsupervised learning in the last decade. Notable examples are deep autoencoders \cite{deep_autoencoder_1}, Restricted Boltzmann Machines (RBMs) \cite{IEEEexample:rbm1}, variational autoencoders \cite{var_auto} and, more recently, Generative Adversarial Networks (GANs) \cite{gan}.
GANs are often used in the context of generative modeling, where they attempt to minimize a measure of discrepancy between a distribution generated by a neural network and the data distribution.
Beyond generative modeling, some works have extended this framework to learn features that are invariant to different domains \cite{ganin} or to noise conditions \cite{dima_invariant}.  Moreover, we recently witnessed some remarkable attempts to learn unsupervised representations by minimizing or maximizing Mutual Information (MI) \cite{philemon,mine,cpc_deepmind,devon_infomax}.
This measure is a fundamental quantity for estimating the statistical dependence between random variables and is defined as the Kullback-Leibler (KL) divergence between the joint distribution over these random variables and the product of their marginal distributions \cite{applebaum_book}. 
As opposed to other metrics, such as correlation, MI can capture complex non-linear relationships between random variables \cite{mi_ref}. MI, however, is difficult to compute directly, especially in high dimensional spaces \cite{mi_comp}. The aforementioned works found that it is possible to maximize or minimize the MI within a framework that closely resembles that of GANs. Additionally, \cite{mine} has further proved that it is even possible to explicitly compute it by exploiting its \textit{Donsker-Varadhan} bound. 

Here we attempt to learn good speaker representations by maximizing the mutual information between two encoded random chunks of speech sampled from the same sentence. 
Our architecture employs both an encoder, that transforms raw speech samples into a compact feature vector, and a discriminator. The latter is alternatively fed by samples from the joint distribution (i.e. two local encoded vectors randomly drawn from the same speech  sentence) and from the product of marginal distributions (i.e, two local encoder vectors coming different utterances). The discriminator is jointly trained with the encoder to maximize the separability of the two distributions. We called our approach \textit{Local Info Max (LIM)} to highlight the fact that it is simply based on randomly sampled local speech chunks. 
Our encoder is based on SincNet \cite{SincNet,sincnet_irasl}, which efficiently processes the raw input waveforms with learnable band-pass filters based on sinc functions. 

The experimental results show that our approach learns useful speaker features, leading to promising results on speaker identification and verification tasks. Our experiments are conducted in both unsupervised and semi-supervised settings and compare different objective functions for the discriminator. We release the code of LIM  within the PyTorch-Kaldi toolkit \cite{pytorch_kaldi}. 

\section{Speaker Representations based on MI}
\label{sec:arch}
The mutual information between two random variables $z_{1}$ and $z_{2}$ is defined as follows:

\begin{equation} \label{eq:mi}
\begin{split}
MI(z_1,z_2) & = \int_{z_1}\int_{z_2} p(z_1,z_2) log\bigg(\frac{p(z_1,z_2)}{p(z_1)p(z_2)}\bigg) dz_1dz_2 \\
 & = D_{KL}\big(p(z_1,z_2) || p(z_1)p(z_2) \big),
\end{split}
\end{equation}
where $D_{KL}$ is the Kullback-Leibler (KL) divergence between the joint distribution $p(z_1,z_2)$ and the product of marginals $p(z_1)p(z_2)$. The MI is minimized when the random variables $z_1$ and $z_2$ are statistically independent (i.e., the joint distribution is equal to the product of marginals) and is maximized when the two variables contain the same information (in which case the MI is simply the entropy of any one of the variables).

 \begin{figure}[t!]
 \centering
   \includegraphics[scale=0.55]{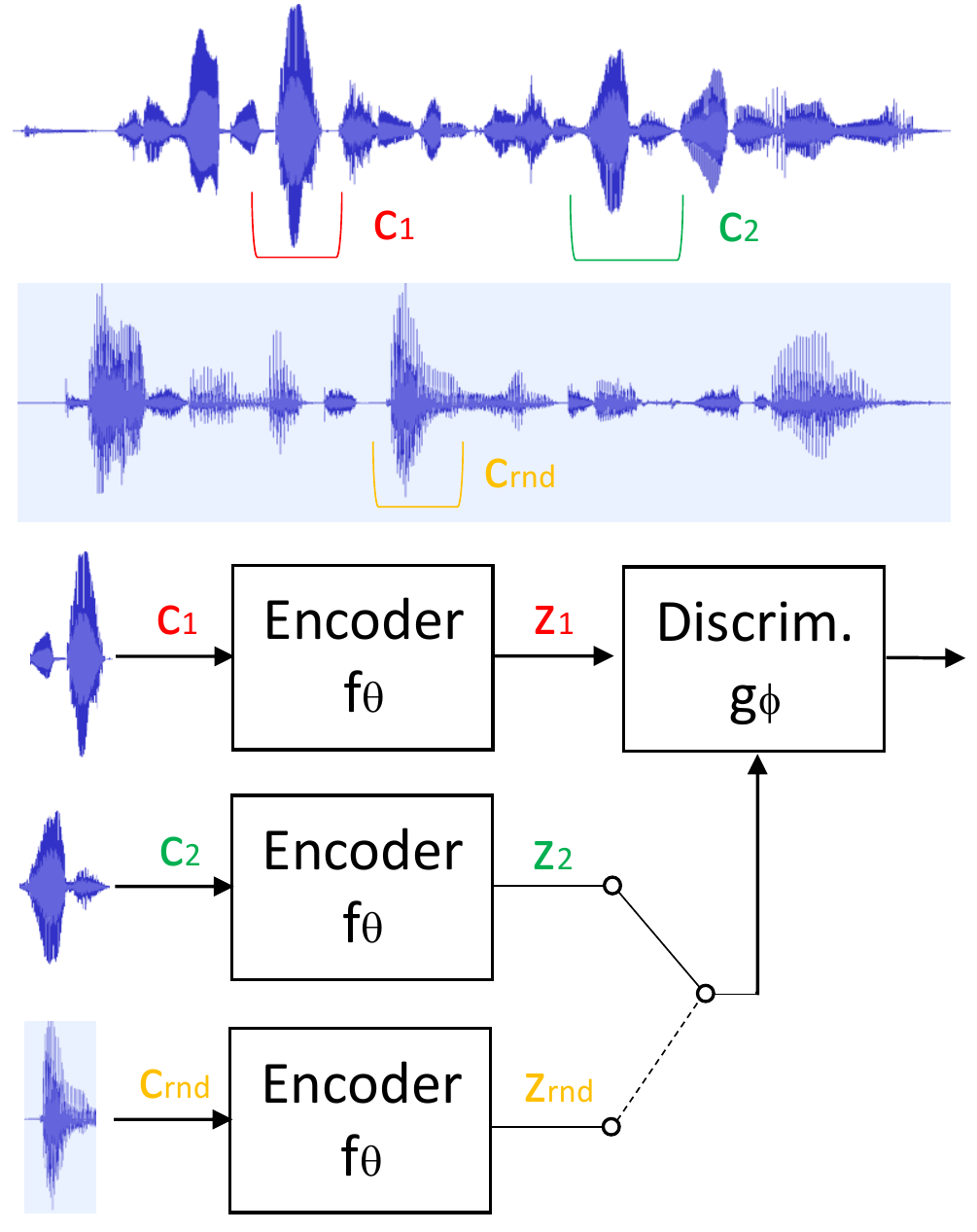}
 \caption{Architecture of the proposed system for unsupervised learning of speaker representations. The speech chunks $c_1$ and $c_2$ are sampled from the same sentence, while $c_{rand}$ is sampled from a different utterance.}
 \label{fig:arch}
 \end{figure}
 
Our LIM system, depicted in Fig.\ref{fig:arch}, aims to derive a compact representation $z$. The encoder $f_{\Theta}$,  with $f:\mathbb{R}^N \rightarrow 	\mathbb{R}^M$, is fed by N speech samples and outputs a vector composed of M real values, while the discriminator  $g_{\Phi}$, with $g:\mathbb{R}^{2M} \rightarrow \mathbb{R}$, is fed by two speaker representations and outputs a real scalar. 
We learn the parameters  $\Theta$ and $\Phi$ of the encoder and the discriminator such that we maximize the mutual information $MI(z_1,z_2)$:
\begin{equation} \label{eq:max}
(\hat{\Theta},\hat{\Phi})= \argmax{\Theta,\Phi}{MI(z_1,z_2)},
\end{equation}
where the two representations $z_1$ and $z_2$ are obtained by encoding the speech chunks $c_1$ and $c_2$ that are randomly sampled from the same sentence.
Note that one reliable factor that is shared across chunks within each utterance is the speaker identity. The maximization of $MI(z_1,z_2)$ should thus be able to properly disentangle this constant factor from the other variables (e.g., phonemes) that characterize the speech signal but
are not shared across chunks of the same utterance.

As shown in Alg. \ref{alg:alg1}, the maximization of MI relies on a sampling strategy that draws positive and negative samples from the joint and the product of marginal distributions, respectively. As discussed so far, the positive samples $(z_1,z_2)$ are simply derived by randomly sampling speech chunks from the same sentence. Negative samples $(z_1,z_{rnd})$, instead, are obtained by randomly sampling from another utterance. The underlying assumptions considered here are the following: (1) two random utterances likely belong to different speakers, (2) each sentence contains a single speaker only. Under these assumptions, that naturally hold in most of the available speech datasets, our method can be regarded as unsupervised (or, more precisely, self-supervised) because no speaker labels are explicitly used. 

A set of $N_{samp}$ positive and negative examples is sampled to form a minibatch $X=\{X_p,X_n\}$. The minibatch $X$ feeds the discriminator $g_{\Phi}$, that is jointly trained with the encoder. Given $z_1$, the discriminator $g_{\Phi}$ has to decide whether its other input ($z_{2}$ or $z_{rnd}$) comes from the same sentence or from a different one (and generally a different speaker). Differently to the GAN framework, the encoder and the discriminator are not adversarial here but must cooperate to maximize the discrepancy between the joint and the product of marginal distributions. 
In other words, we play a \textit{max-max} game rather than a \textit{min-max} one,  making it easier to monitor the progress of training (compared to GAN training), simply as the average loss of the discriminator.

\begin{algorithm}[t!]
\caption{Learning speaker representation with MI}\label{euclid}
\begin{algorithmic}[1]
\While {Not Converged}
      \For{\texttt{i=1 to $N_{samp}$}}
        \State Sample a chunk $c_{1}$ from a random utterance.
        \State Sample another chunk $c_{2}$ from the same utterance.
        \State Sample a chunk $c_{rnd}$ from another utterance.
        \State Process the chunks with the encoder: 
        \State $z_{1}=f_{\Theta}(c_{1})$, $z_{2}=f_{\Theta}(c_{2})$, $z_{rnd}=f_{\Theta}(c_{rnd})$.
        \State Create positive and negative samples:
        \State $X_{p}[i]$=($z_{1}$,$z_{2}$), $X_{n}[i]$=($z_{1}$,$z_{rnd}$). 
      \EndFor
\State Compute discriminator outputs: $g(X_{p})$, $g(X_{n})$. 
\State Compute Loss $L(\Theta,\Phi)$.
\State Compute Gradients $\frac{\partial L}{\partial \Theta}$, $\frac{\partial L}{\partial \Phi}$. 
\State Update $\Theta$ and $\Phi$ to maximize L.

\EndWhile
\end{algorithmic}
\label{alg:alg1}
\end{algorithm}

Different objectives functions can be used for the discriminator. The simplest solution, adopted in \cite{philemon},\cite{devon_infomax} and \cite{petar}, consists in using the standard binary cross-entropy (BCE) loss\footnote{The output layer must be based on a sigmoid when using BCE.}:

\begin{equation} \label{eq:ce}
L(\Theta,\Phi)= \mathbb{E}_{X_p}[\log(g(z_1,z_2))] + \mathbb{E}_{X_n}[\log(1-g(z_1,z_{rnd}))],
\end{equation}
where $\mathbb{E}_{X_p}$ and $\mathbb{E}_{X_n}$ denote the expectation over positive and negative samples, respectively.  
Such a metric estimates the \textit{Jensen-Shannon} divergence between two distributions rather than the KL divergence. 
Consequently, this loss does not optimize the exact KL-based definition of MI, but a similar divergence between two distributions. Differently from standard MI, this metric is bounded (i.e., its maximum is zero), making the convergence of the architecture more numerically stable. 

As an alternative, it is possible to directly optimize the MI with the MINE objective \cite{mine}:

\begin{equation} \label{eq:mine}
L(\Theta,\Phi)= \mathbb{E}_{X_p}[g(z_1,z_2)] -log\Big( \mathbb{E}_{X_n}[e^{g(z_1,z_{rnd})}]\Big).
\end{equation}
MINE explicitly computes MI by exploiting a lower-bound based on the \textit{Donsker-Varadhan} representation of the KL divergence. 
The third alternative explored in this work is the Noise Contrastive Estimation (NCE) loss proposed in \cite{cpc_deepmind}, that is defined as follows:

\begin{equation} \label{eq:nce}
L(\Theta,\Phi)= \mathbb{E}_{x}\bigg[g(z_1,z_2)-\log\bigg(g(z_1,z_2)+\sum_{x_{n}}e^{g(z_1,z_{rnd})}\bigg)\bigg],
\end{equation}
where the minibatch $X$ is composed of a single positive sample and $N-1$ negative samples.
In \cite{cpc_deepmind} it is proved that maximizing this loss maximizes a lower bound on MI.

All the aforementioned objectives are based on the idea of maximizing a discrepancy between the joint and product of marginal distributions. Nevertheless, such losses might be more or less easy to optimize within the proposed framework. 


The unsupervised representations $z$ are then used to train a speaker-id classifier in a standard supervised way. Beyond unsupervised learning, this paper explores two semi-supervised variations for learning speaker representations. The first one is based on pre-training the encoder with the unsupervised parameters and fine-tuning it together with the speaker-id classifier. As an alternative,  we jointly train encoder, discriminator, and speaker-id networks from scratch. This way, the gradient computed within the encoder not only depends on the supervised loss but also on the unsupervised objective. The latter approach turned out to be very effective, since the unsupervised gradient acts as a powerful regularizer.

Similarly to  \cite{palaz_raw,tara_raw,tuske,dnn_emotion,raw_speaker_id}, we propose to directly process raw waveforms rather than using standard MFCC, or FBANK features. The latter hand-crafted features are originally designed from perceptual evidence and there are no guarantees that such inputs are optimal for all speech-related tasks. Standard features, in fact, smooth the speech spectrum, possibly hindering the extraction of crucial narrow-band speaker characteristics such as pitch and formants, that are important clues on the speaker identity. To better process raw audio, the encoder is based on \textit{SincNet} \cite{SincNet,sincnet_irasl}, a novel Convolutional Neural Network (CNN) that encourages the first layer to discover more meaningful filters. 
In contrast to standard CNNs, which learn all the elements of each filter, only low and high cutoff frequencies of band-pass sinc-based filters are directly learned from data, making SincNet suitable to process the high-dimensional audio. 

\section{Related Work}
\label{sec:rel_work}
Similarly to this work, other attempts have recently been made to learn unsupervised representations with mutual information. 
In \cite{philemon}, a GAN that minimizes MI using positive and negative samples has been proposed for Independent Component Analysis (ICA).
A similar approach can be used to maximize MI. In \cite{cpc_deepmind} authors proposed a method called Contrastive Predicting Coding (CPC), that learns representations by predicting the future in a latent space. It uses an autoregressive model optimized with a probabilistic contrastive loss. In \cite{devon_infomax} authors introduced DeepInfoMax (DIM), an architecture that learns representations based on both local and high-level global information. 


The proposed LIM differs from the aforementioned works in the following way: DIM performs a maximization of MI between local and global representations, CPC relies on future predictions, while our method is simply based on random local sampling. Note that training using local embeddings only is very efficient since it does not require the expensive computation of a global representation as in GIM. 
LIM is also related with the recently-proposed methods based on triplet loss \cite{triplet_loss, triplet_loss_spk_rec}. Most of the previous works on triplet loss  (with the exception of \cite{unsup_spk_id}) rely on the speaker labels \cite{triplet_loss_spk_rec}. Moreover, they simply maximize the Euclidean or cosine distance between speaker embeddings. LIM, instead, is based on maximizing the mutual information, thus considering a more meaningful divergence that can also capture complex non-linear relationships between the variables. Maximum Mutual Information (MMI) is often used in HMM-DNN speech recognition as a loss function \cite{MMI}. This loss maximizes the MI between the acoustic probabilities and the targeted word sequence in a standard supervised framework, while LIM is used in a totally different unsupervised context that relies on local speech embeddings. 
Our work also uses SincNet \cite{SincNet,sincnet_irasl} (that is here used for the first time in an unsupervised framework), and extends the previous works by also addressing semi-supervised learning where encoder, discriminator, and speaker-id classifier are jointly trained from scratch. Moreover, to the best of our knowledge, this paper is the first that compares several objective functions for MI optimization in a speech task.

\section{Experimental Setup}\label{sec:setup}
The proposed method has been evaluated using different corpora. 
In the following, an overview of the experimental setting is provided.

\subsection{Corpora}
This paper considered the TIMIT (462 spks, \textit{train} chunk)  \cite{timit}, Librispeech  (2484 spks), and VoxCeleb1 (1251 spks) \cite{voxceleb}  corpora. 
To make TIMIT and Librispeech speaker recognition tasks more challenging, we only employed 12-15 seconds of randomly selected training material for each speaker. Moreover, a set of TIMIT and Librispeech experiments have also been performed in distant-talking reverberant conditions. In this case, all the clean signals were convoluted with a different impulse response, that was sampled from the DIRHA dataset  \cite{dirha_asru,ravanelli}. The DIRHA corpus contains high-quality multi-room and multi-microphone impulse responses, that were measured in a domestic environment with a considerable reverberation time of $T_{60}=0.7 s$. This way, we are able to provide experimental evidence in a much more challenging acoustic scenario and we can introduce a channel effect that is not natively present in the clean TIMIT and Librispeech corpora. 
To study our approach using a more standard speaker recognition dataset, we also employed the VoxCeleb1 corpus (using the provided lists).


\subsection{DNN Setup}
The waveform of each speech sentence was split into chunks of 200 ms (with 10 ms overlap), which were fed into the SincNet encoder. The first layer of the encoder performs sinc-based convolutions, using 80 filters of length $L=251$ samples. The architecture then employs two standard convolutional layers, both using 60 filters of length 5. Layer normalization \cite{layer_norm} was used for both the input samples and for all convolutional layers. Next, two fully-connected leaky-ReLU layers \cite{leaky_relu}  composed of 2048 and 1024 neurons (normalized with batch normalization \cite{batchnorm,ravanelli_SLT}) were applied. 
Both the discriminator and the speaker-id classifier are fed by the encoder output and consist of MLPs based on a single ReLU layer. Frame-level speaker classification was obtained from the speaker-id network by applying a softmax output layer, that provides a set of posterior probabilities over the targeted speakers. A sentence-level classification was derived by averaging the frame predictions and voting for the speaker which maximizes the average posterior.
Training used the RMSprop optimizer, with a learning rate $lr=0.001$, $\alpha=0.95$, $\epsilon=10^-7$, and minibatches of size 128. 
All the hyper-parameters of the architecture were tuned on TIMIT, then inherited for Librispeech and VoxCeleb as well.  

The speaker verification system was derived from the speaker-id neural network using the \textit{d-vector} technique. The \textit{d-vector} \cite{dnn_spk_rec_class2,voxceleb} was extracted from the last hidden layer of the speaker-id network. A speaker-dependent d-vector was computed and stored for each enrollment speaker by performing an L2 normalization and averaging all the d-vectors of the different speech chunks. The cosine  distance between enrolment and test d-vectors was then calculated, and a threshold was then applied on it to reject or accept the speaker.
Note that to assess our approach on a standard open-set speaker verification task, all the enrolment and test utterances were taken from a speaker pool different from that used for training the speaker-id DNN.

\section{Results}
\label{sec:res}
This section summarizes our experimental activity on speaker identification and verification. 

\subsection{Speaker Identification}
Tab. \ref{tab:unsup} reports the sentence-level classification error rates achieved with binary cross-entropy (BCE), MINE,  Noise Constructive Estimation (NCE), and the triplet loss used in \cite{triplet_loss_spk_rec}. 

\begin{table}[h]
\centering
\setlength{\tabcolsep}{4pt}
\begin{tabular}{l|cc|cc}  
& \multicolumn{2}{|c|}{TIMIT} & \multicolumn{2}{c}{Librispeech}  \\ 
& CNN & SincNet & CNN & SincNet \\ \cline{1-5}
Unsupervised-Trip. Loss & 2.84 & 2.22 & 1.46 & 1.33 \\
Unsupervised-MINE & 2.15 & 1.36 & 1.43 & 0.94 \\
Unsupervised-NCE & 2.05 & 1.29 & 1.14 & 0.82 \\
Unsupervised-BCE & 1.98 & \textbf{1.21} & 1.12 & \textbf{0.75} \\
\bottomrule
\end{tabular}
\caption{Classification Error Rate (CER\%) obtained on TIMIT (462 spks) and Librispeech (2484 spks) speaker-id tasks using LIM embeddings learned with various objective functions.}
\label{tab:unsup}
\end{table}

The table highlights that our LIM embeddings contain information on the speaker identity, leading to a CER(\%) ranging from  2.84\% to 1.21\% in all the considered settings.
It is worth noting that mutual information losses (i.e., MINE, NCE, BCE) outperform the triplet loss.
This result suggests that better embeddings can be derived with a divergence measure more meaningful than the simple cosine distance.
The best performance is achieved with the standard binary cross-entropy. Similar to \cite{devon_infomax}, we have observed that this bounded metric is more stable and more easy to optimize. Both MINE and NCE objective are unbounded and their value can grow indefinitely during training, eventually causing numerical issues. 
The performance achieved with Librispeech is better than that observed for TIMIT. Even though the former is based on more speakers, its utterances are on average longer than the TIMIT ones.
The table also shows that SincNet outperforms a standard CNN. This confirms the promising achievements obtained in \cite{SincNet,sincnet_irasl} in a standard supervised setting. SincNet, in fact, converges faster and to a better solution, thanks to the compact sinc filters that make learning from high-dimensional raw samples easier.  

Tab. \ref{tab:spk_id} extends previous speaker-id results to other training modalities, including supervised and semi-supervised learning in both clean and reverberant acoustic conditions.
\begin{table}[h]
\centering
\begin{tabular}{l|cc|cc}  
& \multicolumn{2}{|c|}{TIMIT} & \multicolumn{2}{c}{Librispeech}  \\ 
& Clean & Rev & Clean & Rev \\ \cline{1-5}
Supervised & 0.85 & 34.8 & 0.80 & 17.1 \\
Unsupervised-BCE & 1.21 & 28.2 & 0.75 & 15.2 \\
Semi-supervised-pretr. & 0.69 & 25.4 & 0.56 & 9.6 \\
Semi-supervised-joint & \textbf{0.65} & \textbf{24.6} & \textbf{0.52} & \textbf{9.3} \\
\bottomrule
\end{tabular}
\caption{Classification Error Rate (CER\%) obtained on speaker-id with supervised, unsupervised and semi-supervised modalities in clean and reverberat conditions.}
\label{tab:spk_id}
\end{table}

From the table, it emerges that the results achieved when feeding the classifier with our speaker embeddings (\textit{unsupervised-BCE}) are often better than those obtained with the standard supervised training (\textit{supervised}). 
The gap becomes more evident when we pass from unsupervised to semi-supervised learning. In particular, the joint semi-supervised framework (i.e., the approach that jointly trains encoder, discriminator, and speaker classification for scratch) yields the best performance, surpassing the performance obtained when pretraining the encoder and then fine-tuning it with the supervised task (\textit{Semi-supervised-pretr.}). The internal representations discovered in this way are influenced by both the supervised and the unsupervised loss. The latter one acts as a powerful regularizer, that allows the neural network to find robust features. 
The results also show a significant performance degradation in distant-talking acoustic conditions. The presence of considerable reverberation and the introduction of channel/microphone variabilities, in fact, make speaker-id particularly challenging.

\subsection{Speaker Verification}
We finally extend our validation to speaker verification on the VoxCeleb corpus. Table \ref{tab:eer} compares the Equal Error Rate (EER\%) achieved using our best system (\textit{Semi-supervised-pretr.}) with some previous works on the same dataset.
\begin{table}[h]
\centering
\begin{tabular}{l|c}  
& EER (\%) \\ \hline
GMM-UBM \cite{voxceleb}  & 15.0  \\
I-vectors + PLDA \cite{voxceleb}  & 8.8  \\
CNN \cite{voxceleb}  & 7.8  \\
CNN + intra-class + triplet loss\cite{voxceleb_res}  & 7.9  \\ 
SincNet \cite{SincNet}  & 7.2 \\
SincNet+LIM (proposed) & \textbf{5.8} \\ 
\bottomrule
\end{tabular}
\caption{Equal Error Rate (EER\%) obtained on speaker verification (using the VoxCeleb corpus).}
\label{tab:eer}
\end{table}

The proposed model reaches an EER(\%) of 5.8\% and outperforms other systems such as an I-vector baseline \cite{voxceleb,i-vect_short}, a standard CNN \cite{voxceleb}, and a CNN based on combination of intra-class and triples loss \cite{voxceleb_res}. Finally, LIM outperforms a standard SincNet model trained in a fully supervised way \cite{SincNet}. This result  confirms the effectiveness of the proposed approach even in an open-set text-independent speaker verification setting.  

\section{Conclusion}
\label{sec:conc}
This paper proposed a method for learning speaker embeddings by maximizing mutual information. The experiments have shown promising performance on speaker recognition and have highlighted better results when adopting the standard binary cross-entropy loss, that turned out to be more stable and easier to optimize than other metrics. It also highlighted the importance of using SincNet, confirming its effectiveness when processing raw audio waveforms. The best results are obtained with end-to-end semi-supervised learning, where an ecosystem of neural networks composed of an encoder, a discriminator, and a speaker-id must cooperate to derive good speaker embeddings.
Our achievement can be easily combined with other recent findings in speaker recognition. For instance, it is possible to use LIM to extract semi-supervised x-vectors. We can also improve it by employing an attention mechanism that weights the contribution of each time frame, or by combing our semi-supervised costs with other losses, such as the center loss.


\section{Acknowledgment}
We would like to thank D. Hjelm, T. Parcollet, and M. Omologo, for helpful comments.
This research was enabled by support provided by Calcul Qu\'ebec and Compute Canada.

\bibliographystyle{IEEEtran}

\bibliography{mybib}


\end{document}